\documentclass[reprint,amsmath,amssymb,aps,pra,longbibliography]{revtex4-2}
\usepackage{algorithmic}
\usepackage{graphicx}
\usepackage{textcomp}
\usepackage{lipsum,mathtools,cuted}
\usepackage{float}
\usepackage{epsfig}
\usepackage{epstopdf}
\usepackage{subfigure,color}
\usepackage{dcolumn}
\usepackage{accents}
\usepackage{multirow}
\usepackage[usestackEOL]{stackengine}
\usepackage{tabularx}
\usepackage[normalem]{ulem}
\usepackage{makecell}
\usepackage{dcolumn}
\usepackage{bm}
\usepackage{booktabs} 
\usepackage{comment}
\usepackage{times}
\usepackage{amsfonts}
\usepackage{esvect}
\usepackage{harpoon}
\usepackage{amsmath,accents}
\usepackage{subfigure,color}
\usepackage{graphicx}
\usepackage{dcolumn}
\usepackage{bm}
\usepackage{comment}
\usepackage[mathlines]{lineno}

\newcommand{\ket}[1]{\left \rvert #1 \right \rangle}
\newcommand{\E}[1]{\left \langle #1 \right \rangle}

\begin{document}

\title{Broadband amplification of light through adiabatic spatiotemporal modulation}
\author{M.~H.~Mostafa$^{1}$}
\author{M.~S.~Mirmoosa$^{2}$} 
\author{E. Galiffi$^{3}$}
\author{S. Yin$^{3}$}
\author{A. Alù$^{3,4}$} 
\author{S.~A.~Tretyakov$^{1}$} 
 
\affiliation{$^{1}$Department~of~Electronics~and~Nanoengineering, Aalto~University, P.O.~Box~15500, FI-00076~Aalto, Finland\\$^{2}$Department of Physics and Mathematics, University of Eastern Finland, P.O.~Box~111, FI-80101 Joensuu, Finland\\$^{3}$
Photonics Initiative, Advanced Science Research Center, City University of New York, New York, NY, USA.\\$^{4}$Physics Program, Graduate Center of the City University of New York, New York, NY, USA.}


\begin{abstract}
Four-dimensional optics leverages the simultaneous control of materials in space and time to manipulate light. A key challenge in experimentally realizing many intriguing phenomena is the need for rapid modulation, which is hindered by the inherently adiabatic relaxation of optical materials. Here, we theoretically demonstrate that broadband amplification can be achieved without the need for sub-cycle temporal responses, instead leveraging adiabatic spatiotemporal modulation patterns. The proposed modulation scheme is compatible with recent demonstrations of the temporal modulation of epsilon-near-zero materials. We also show that the same phenomenon may be realized by modulating bianisotropic nonreciprocal media in time. This broadband gain mechanism opens new avenues for the generation of high-energy, ultrashort optical pulses, with potential impact in ultrafast optics and electron microscopy.
\end{abstract}

\maketitle


Modulating the optical properties of materials in time gives rise to various optical phenomena, including amplification~\cite{morgenthaler1958velocity, lyubarov2022amplified}, frequency conversion~\cite{PhysRevA.62.033805, mendoncca2002time, Agrawal2014RTC}, and photon-pair generation~\cite{Aleksandr2005, sajjad:Q}. More specifically, materials and metamaterials characterized by abrupt changes in their optical properties in time, supporting temporal interfaces, have garnered significant interest in recent times~\cite{Alu_TL,jones2024time,Xiao:14, Mostafa:24, ramaccia2020light}. Temporal interfaces produce forward and backward waves, associated with temporal refraction and reflection, and dual to the refraction and reflection observed at spatial interfaces. Unlike spatial interfaces, temporal interfaces do not conserve energy, which opens the possibility of amplification. 

The growing interest around parametric amplification of light through temporal variations in the optical properties of materials~\cite{Emanuele_luminal, lyubarov2022amplified,Ptitcyn2023Tutorial, WaMIR23PTC} has driven research into periodic temporal interfaces, giving rise to photonic time crystals (PTCs) that exhibit momentum bandgaps characterized by imaginary eigenfrequencies, associated with exponentially growing eigenmodes, i.e., parametric gain~\cite{Halevi9PTC,Segev8PTC,lyubarov2022amplified,Boltasseva:24,asgari2024photonic}. The eigenmodes within a momentum bandgap are standing waves formed through temporal reflections, implying that the amplification in PTCs is intrinsically reliant on the access to temporal reflections (photon-pair generation~\cite{PhysRevA.62.033805}), and hence require modulation at sub-cycle timescales. 

In contrast, this work proposes a scheme that achieves gain through adiabatic modulation (not relying on temporal reflections) by incorporating spatial modulations and carefully designing both the spatial and temporal modulation profiles.

Indeed, while PTCs offer intriguing potential, their practical realization at near-optical and optical frequencies remains challenging due to their reliance on temporal reflections. Although significant progress has been made in ultrafast optics, temporal reflections have so far been demonstrated only at radio frequencies~\cite{moussa2023, jones2024}. Achieving temporal reflection requires large impedance switching within a timescale comparable to a single cycle of the input wave~\cite{tapered}. Strong modulation is necessary to generate substantial temporal reflection at each time interface and, consequently, achieve gain. These stringent conditions make the implementation of PTCs in optical regimes particularly demanding, as both speed and modulation strength requirements must be met simultaneously by the modulation mechanism~\cite{Boltasseva:24,Hayran:22}.

While PTCs utilize only the temporal degree of freedom, limiting the scope of functionalities, four-dimensional optics involving spatiotemporal engineering of optical materials have significantly expanded the spectrum of functionalities for wave engineering~\cite{engheta2023four,galiffi2022photonics,Caloz:1,Caloz:2}, with numerous recent experiments exploiting time-varying metasurfaces and slabs~\cite{guo2019nonreciprocal, harwood2024super, zhou2020broadband,WaMIR23PTC,segev:single,tirole2023double, photon_acc, adiabatic_f}. Spatiotemporal modulation refers to the dynamic manipulation of light across both spatial and temporal dimensions, either cascaded or simultaneous, enabling intriguing effects, such as broken reciprocity~\cite{Yakir2015}, Fresnel drag~\cite{Paloma_drag}, light amplification in trans-luminal metamaterials~\cite{Emanuele_luminal,Pendry:21,Pendry_forceline,Pendry:con,Pendry:ph}, cascaded frequency transformations~\cite{Fort22,PachecoPena24}, and more~\cite{PhysRevA.110.043526,liberal:Noether, Wedges, Caloz2024}.

A disconnect persists between theoretical and experimental studies on amplification via spatiotemporal modulation: theoretical models typically assume fast switching, whereas optical and near-optical experimental implementations are typically limited to adiabatic switching. We fill this gap by proposing broadband amplification through cascades of adiabatic spatiotemporal modulations, emphasizing how this method leverages a gain mechanism based on frequency up-conversion~\cite{Yakir2020, Pendry:con, PhysRevA.110.043526, Fort22,liberal:Noether}. In addition, we highlight how current experimental setups already hold the potential to realize this amplification protocol at near-optical frequencies~\cite{guo2019nonreciprocal, harwood2024super, zhou2020broadband,segev:single,tirole2023double, photon_acc, adiabatic_f}.

\begin{figure}[t!]
\centerline
{\includegraphics[width=1\linewidth]{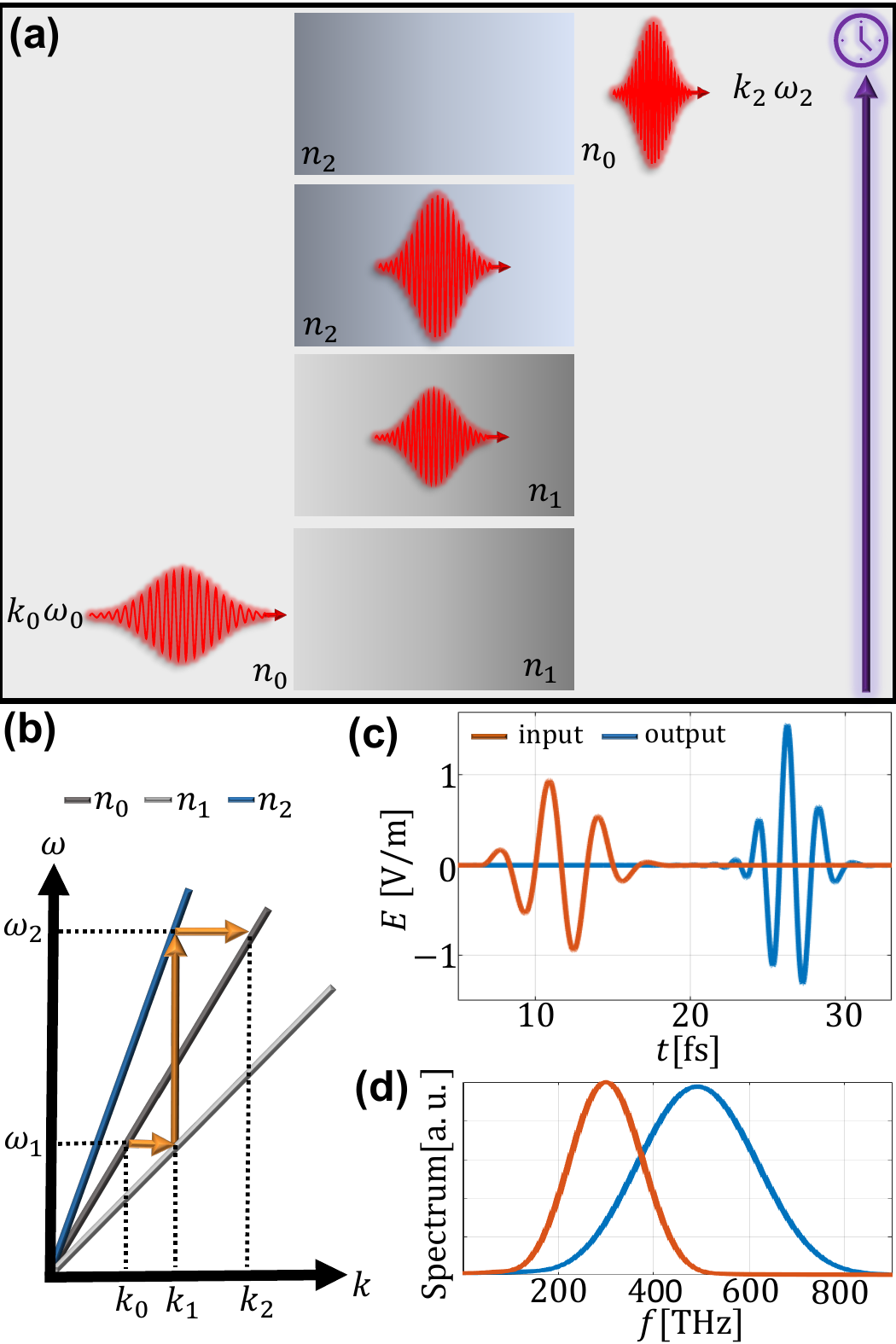}}
\caption{(a) Schematic illustration of a wavepacket traversing a cascade of adiabatic spatial and temporal interfaces, with the time axis oriented vertically. The wavepacket goes through an adiabatic spatial interface where the refractive index changes from $n_0$ to $n_1$, then experiences an adiabatic temporal interface altering the refractive index to $n_2$, and finally exits through another adiabatic spatial interface to propagate in the original medium $n_0$. Note that all interfaces are adiabatic (impedance-matched). (b) Dispersion diagram showing the frequency/energy and wavenumber/momentum transformations. (c)  Electric field of a Gaussian pulse going through the setup in (a), where $n_0=1$, $n_1=1.73$, and $n_2=1.048$.  (d) Spectrum of the fields in (c), where the frequency/energy is transformed correctly with the factor $n_1/n_2$. Simulations are done using COMSOL Multiphysics.}
\label{fig_0}
\end{figure}

\noindent{\textit{Adiabatic spatiotemporal modulation of a slab}}---
Consider a Gaussian pulse with center frequency $\omega_0$ and wavenumber $k_0$, propagating through a cascade of adiabatic (impedance-matched) spatial and temporal interfaces as shown in Fig.~\ref{fig_0}(a). At the first adiabatic spatial interface between materials with refractive indices $n_{0}$ and $n_{1}$, the linear (Minkowski) momentum and the wavenumber are transformed by a factor $n_1/n_0$.  Such spatial interface conserves frequency and energy while inducing no reflections, and has an electric field transmission coefficient of $1$.

Next, an adiabatic temporal modulation of the refractive index $n_{1}\to n_{2}$ takes place, transforming the frequency by a factor $n_1/n_2$. At abrupt temporal interfaces, the average energy density undergoes a transformation expressed as $\frac{u_2}{u_1}= n_1 / 2n_2 (Z_1/Z_2+Z_2/Z_1)$~\cite{Mostafa:24}, where $\frac{u_2}{u_1}$ is the ratio of the average energy density after the interface to the one before, and $Z_1$ ($Z_2$) is the impedance before (after) switching. By contrast, adiabatic temporal modulation, characterized by a transition duration that exceeds the wave period, facilitates impedance matching and suppresses temporal reflections, as demonstrated in prior work~\cite{tapered}. In this regime, the adiabatic modulation imparts the energy scaling $\frac{u_2}{u_1}=n_1 / n_2$, equal to the frequency transformation.

Finally, another adiabatic spatial interface is implemented to couple the pulse back into the initial material, where the momentum and wavenumber transform by a factor $n_0/n_2$. This leads to a total momentum transformation of $(n_1/n_0)(n_0/n_2)=n_1/n_2$. As a result of one full cycle of the spatiotemporal modulation shown in~Fig.~\ref{fig_0}(a), the frequency, energy,  wavenumber, and momentum increase as shown on the dispersion diagram in~Fig.~\ref{fig_0}(b).
This modulation cycle is repeatable, since the pulse can be guided to go through the same setup multiple times. In this case, the total transformation for frequency, energy, wavenumber, and momentum is given by the factor $(n_1/n_2)^r$, where $r$ denotes the number of modulation cycles. Hence, both energy and momentum grow exponentially with the number of modulation cycles.

These adiabatic spatial interfaces can be implemented using gradient-index layers~\cite{gradient}. With each modulation cycle, the wavelength decreases, such that the variation in the gradient-index layer becomes more smooth relative to the wavelength, and the adiabatic condition is more accurately satisfied as the amplification process unfolds. One modulation cycle has been simulated using COMSOL Multiphysics, and the results are presented in~Figs.~\ref{fig_0}(c) and (d), showing a clear frequency up-shift, and an increase in the average energy density. It is important to note that these transformations are inherently broadband, provided that the interfaces remain adiabatic. The proposed modulation scheme operates within current practical limitations on modulation speeds observed in current pump-probe setups at near-optical frequencies~\cite{guo2019nonreciprocal, harwood2024super, zhou2020broadband,segev:single,tirole2023double, photon_acc, adiabatic_f}. 

Crucially, a single slab operates effectively only within a limited frequency range, due to the inherent frequency dispersion of practical materials. Consequently, the dispersion characteristics of the slab determine its operational bandwidth. To extend the frequency coverage, multiple slabs can be cascaded, with each designed to target a distinct frequency band.

\noindent{\textit{Amplification and photon number conservation}}---Recent studies have investigated temporal interfaces in the quantum regime, offering new insights into the associated physical phenomena~\cite{lyubarov2022amplified, sajjad:Q} . A particularly relevant study~\cite{Liberal2023} examines the classical limit by modeling the system as a coherent state with a large average number of photons. The study demonstrates that implementing an antireflection temporal coating enables frequency transformation while conserving the average number of photons, effectively suppressing photon-pair generation. Additionally, it is shown that the Hamiltonian scales proportionally with the frequency, thus permitting energy gain.

In the following, we show that such gain mechanism is not intrinsically dependent on the use of antireflection temporal coatings, which remain difficult to realize at optical frequencies, but rather requires impedance matching generally. 
Consider a single mode, with a given wavevector and polarization state, represented by a coherent state $\ket{\alpha}$. Assuming that the propagation is in the $z$ direction, the coherent field  (i.e., the expectation value of the scalar electric-field operator) is given by~\cite{quantum.optics}
\begin{equation}
\E{\hat{E}_1(z, t)}= 2 \sqrt{\frac{\hbar \omega_1}{2\epsilon _1V}} |\alpha_1| \cos{(\chi_1-\theta_1)},
\label{eq:E}
\end{equation} 
where $\chi_1=\omega_1t-k_1z-\pi/2$, $|\alpha_1|$ is the square root of the average number of photons, $\hbar$ is the reduced Planck constant, $\epsilon_1$ is the permittivity of the host medium, $V$ is the quantization volume, $\omega_1$ is the angular frequency, $k_1$ is the wavenumber, and $\theta_1$ is the phase related to the coherent complex parameter $\alpha_1$. The expectation value of the Hamiltonian reads $\E{\hat{H}_1}\approx \hbar \omega_1 |\alpha_1|^2$ where the average number of photons $|\alpha_1|^2$ is assumed to be much larger than one. Upon a temporal interface in linear, dispersionless, lossless and spatially homogeneous materials, the well-known temporal boundary conditions apply, being the continuity of the flux density operators $\hat{{D}}_1(t=0^-)=\hat{{D}}_2(t=0^+)$ and $\hat{{B}}_1(t=0^-)=\hat{{B}}_2(t=0^+)$~\cite{morgenthaler1958velocity, PhysRevA.62.033805, sajjad:Q}. The boundary conditions allow a discontinuity in $\E{\hat{E}}$ at the temporal interface, indicating a change in the amplitude $|\alpha|$. By computing the shift in $\omega$ using the dispersion relation, the amplitude  $|\alpha|$ remains the sole unknown. It can then be determined by solving the boundary condition equations  as shown in~\cite{SM}. We express the temporal refraction $T$ and temporal reflection $R$ coefficients with respect to the average number of photons $|\alpha_1|^2$ as
\begin{align}
T= \frac{1}{4} \left(\sqrt{\frac{Z_2}{Z_1}}+\sqrt{\frac{Z_1}{Z_2}} \right)^2,\\
R= \frac{1}{4} \left(\sqrt{\frac{Z_2}{Z_1}}-\sqrt{\frac{Z_1}{Z_2}}\right)^2.
\end{align}
The temporal reflection coefficient $R$ quantifies the average number of photon pairs produced at the temporal interface, as an equal amount is produced in the forward and backward directions, to satisfy the (Minkowski) momentum conservation relation $T-R=1$~\cite{tapered}. Under the assumption $|\alpha_1|^2\gg1$ in which the photons produced by vacuum amplification are negligible~\cite{Liberal2023}, the expectation value of the Hamiltonian after the temporal interface reads $\E{\hat{H_2}} \approx \hbar \omega_1 \frac{n_1}{n_2} (T+R)|\alpha_1|^2$, where the factor $n_1/n_2$ indicates the frequency conversion that results from index variation under conservation of the individual photon momentum $n_1\omega_1=n_2\omega_2$. For an impedance matched temporal interface, $R=0$ and $T=1$. However, the energy is still transformed by the factor $\frac{n_1}{n_2}$ due to the frequency conversion. Antireflection temporal coatings is not the only possible way to ensure impedance matching. As shown in \cite{tapered}, it can be realized through adiabatic switching.

This theory of Hamiltonian translation explains how the adiabatic temporal modulation in Fig.~\ref{fig_0}(a) enables broadband amplification of photon energy. The process relies solely on frequency translation, occurring under adiabatic switching that is much slower than a single optical cycle~\cite{Yakir2020}, or with a suitably tailored temporal profile~\cite{tapered}.

This semi-classical formulation highlights how energy is exchanged between light and the mechanism responsible for the temporal modulation, distinguishing two mechanisms through which light amplification occurs due to time modulations. The first possible mechanism exploits photon-pair generation, achieving gain through increasing the total average number of photons $|\alpha|^2$~\cite{lyubarov2022amplified, sajjad:Q}, which relies on rapid switching of the wave impedance $Z=\sqrt{\frac{\mu}{\epsilon}}$. The second gain mechanism is based on increasing the energy of each photon $\hbar \omega$ through frequency translation: by decreasing the refractive index ($n_1>n_2$), the energy of individual photons is elevated to $\hbar\omega_1 \frac{n_1}{n_2}>\hbar\omega_1$ while the average number of photons remains the same~\cite{Yakir2020, Pendry:con, PhysRevA.110.043526, Fort22}. This second mechanism is also present in energy exchanges upon collisions between photons and relativistic charged particles (mainly electrons) in high-energy physics and astrophysics~\cite{longair2011high}.

\begin{figure}[b!]
\centerline
{\includegraphics[width=1\linewidth]{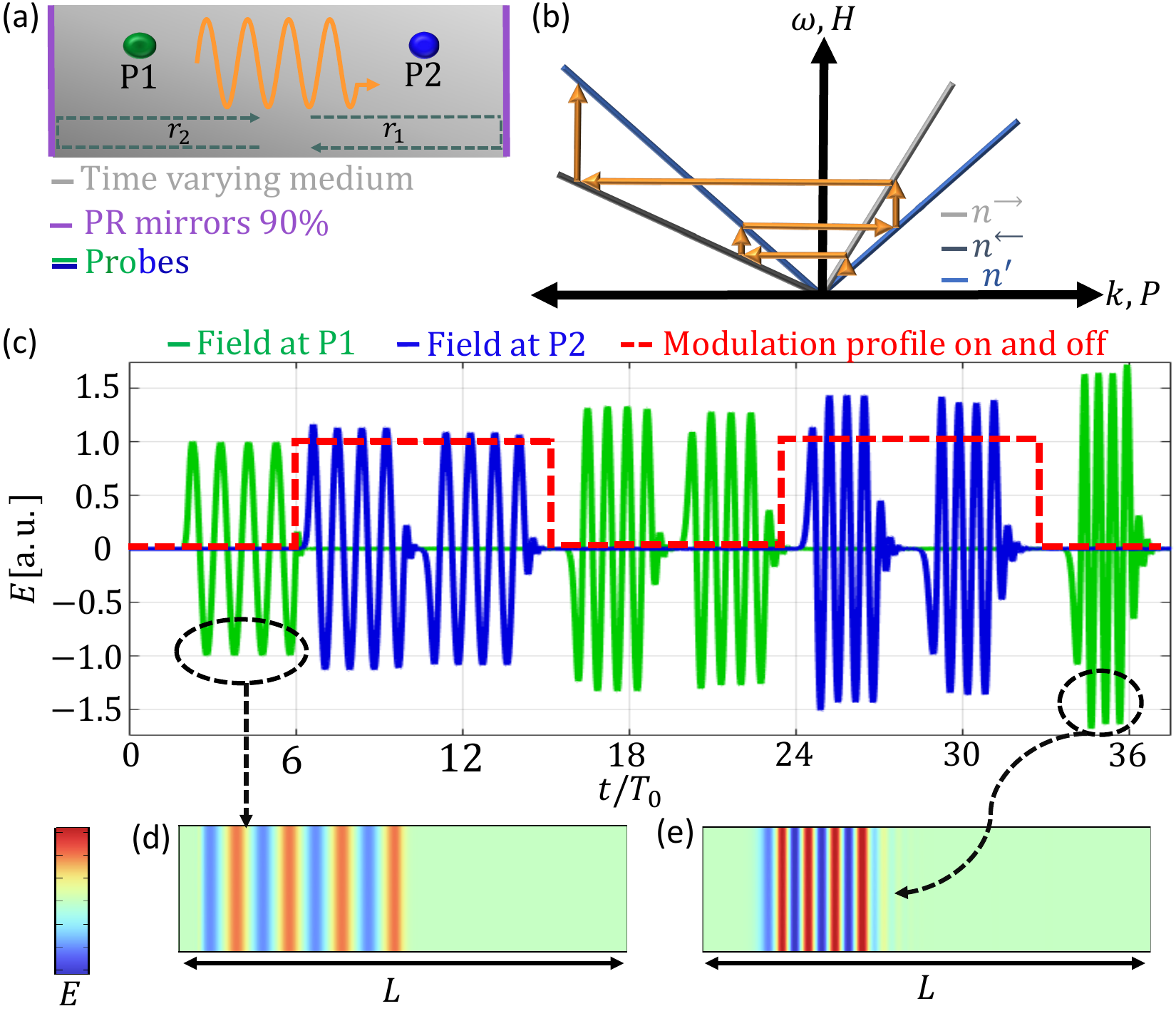}}
\caption{ (a) Schematic illustration of a cavity (mirrors' reflectance $\approx0.90$) filled with a time-varying NBM. (b) Schematic of the dispersion diagram showing the cascaded up-conversion of frequency, energy, wavenumber and momentum taking place inside the cavity. (c) Temporal evolution of the electric field inside the cavity while the NBM is switched periodically on and off, where $\epsilon'=2\epsilon_0$, $\mu'=\mu_0$, $\frac{\epsilon^\leftarrow}{\epsilon'}=\frac{\mu^\leftarrow}{\mu'}=1.23$ and $\frac{\epsilon^\rightarrow}{\epsilon'}=\frac{\mu^\rightarrow}{\mu'}=0.88$. The time axis is normalized to the period of the initial wave $T_0$. With every switching event, the frequency is up-converted and the field is amplified. (d)-(e) Spatial profile of the initial and final electric field, respectively. The wavelength is clearly shifted down, as the wavenumber (momentum) is up-converted.} 
\label{fig_2}
\end{figure} 

\noindent{\textit{Photon number-conserving amplification with bianisotropic temporal interfaces}}---
The described amplification mechanism relies on shifting both frequency and wavenumber to larger values, as shown in Fig.~\ref{fig_0}(b). In this section, we show that, by exploring this phenomenon in nonreciprocal bianisotropic (artificially moving) materials (NBMs)~\cite{Sajjad2024, Mostafa:24}, we may eliminate the need for spatial interfaces, leveraging the direction-dependent phenomena inherent to NBMs and their nonreciprocal response. The flux density associated with NBM are given by~\cite{Sajjad2024}
\begin{equation}
{D}^\rightleftarrows=\epsilon^\rightleftarrows {E},~~{B}^\rightleftarrows=\mu^\rightleftarrows {H},
\label{eq:CRMR}
\end{equation} 
in which $\rightleftarrows$ indicate the propagation direction of waves being along $z$ or $-z$ direction. As a result of nonreciprocity, the medium's response is dependent on the propagation direction. The constitutive relations in Eq.~\eqref{eq:CRMR} can be viewed as describing two different isotropic and spatially homogeneous magnetodielectric media, where the propagating field interacts with one of these media depending on its propagation direction. Consequently, the wavenumbers and impedances are $k^\rightleftarrows=\omega\sqrt{\epsilon^\rightleftarrows \mu^\rightleftarrows}$ and $Z^\rightleftarrows=\sqrt{\mu^\rightleftarrows/\epsilon^\rightleftarrows}$, respectively. We consider temporal interfaces between  NBM and an isotropic stationary magnetodielectric with permitivity $\epsilon'$, permeability $\mu'$, and refractive index $n'$ for the case where interfaces are impedance-matched. The conservation of impedance $Z^\rightleftarrows=\sqrt{\mu^\rightleftarrows/\epsilon^\rightleftarrows}=\sqrt{\mu'/ \epsilon'}$ results in zero temporal reflection, and only temporal refraction is induced. Impedance matching does not require modulation of $\mu$, since it can be achieved through adiabatic switching~\cite{Yakir2020, tapered}. At the temporal interface, the scaling coefficient for energy and frequency reads
\begin{equation}
C^\rightleftarrows_{\text{off}}=\frac{1}{C^\rightleftarrows_{\text{on}}}=\frac{n^\rightleftarrows}{n'},
\label{eq:CoeffREF}
\end{equation} 
where 'on' refers to switching from the isotropic medium to the bianisotropic medium, and 'off' denotes the reverse transition. The scaling of the electric field is equal to the one for energy in Eq.~\eqref{eq:CoeffREF}, since the wave impedance is matched.

The amplification mechanism described above for dielectric media requires the use of both temporal and spatial interfaces. We can implement an effective spatial interface by using a mirror that reverts the propagation direction in an NBM, creating a transition between $n^\rightarrow$ and $n^\leftarrow$. Assuming an initial field propagating along the +$z$ direction, the sequence of cascaded interfaces starts with a temporal interface between $n'$ and $n^\rightarrow$, followed by spatial reflection from a mirror to couple the fields spatially from $n^\rightarrow$ to $n^\leftarrow$, and finally, a second temporal interface between $n^\leftarrow$ and $n'$. Such sequence is illustrated through the round trip $r_1$ in Fig.~\ref{fig_2}(a).  A second round trip, $r_2$, does not involve any transformations. Instead, it is utilized to couple the field back to its initial position, resetting the system to its initial state. In the presented example, the mirrors have 0.9 reflectance to represent losses in the system, while the length of the cavity $L$ is considered to be larger than the total length of the wave packet under consideration. After one complete modulation cycle, the energy, momentum, frequency, wavenumber, and electric field are scaled by the factor $\frac{n^\leftarrow}{n^\rightarrow}$.  In this setup, cascaded scaling take place at each reflection, as illustrated in the dispersion diagram in Fig.~\ref{fig_2}(b). The temporal evolution of the electric field inside the cavity is shown in Fig.~\ref{fig_2}(c), where the field is transformed every time the medium is switched. The total transformation coefficient for the frequency and wavenumber follow $(\frac{n^\leftarrow}{n^\rightarrow})^r = 1.4^2\approx 1.96$, where $r$ is the number of modulation cycles. The total scaling coefficient for the electric field is $(\frac{n^\leftarrow}{n^\rightarrow})^r (\sqrt{0.9})^b = 1.96 \sqrt{0.9}^3 \approx1.67$, and for the total energy it is $(\frac{n^\leftarrow}{n^\rightarrow})^r 0.9^b= 1.96 \cdot 0.9^3 \approx1.42$, where $b$ is the number of reflections from the cavity mirrors. The spatial profile of the initial and final electric fields are shown in Figs.~\ref{fig_2}(d)-(e). The wavelength is clearly shifted down, as the wavenumber (momentum) is up-converted. Figures~\ref{fig_2}(c)-(e) show simulations of cascaded energy and momentum transformations, confirming the theoretical predictions.

The confinement of the field inside the cavity offers several advantages. There is no need for fast modulation, as the transformations take place every time a switching event occurs, independent of the frequency of switching events. As a result, assuming that the switching ON took place at moment $t_\text{on}$, the switching OFF can take place at any later moment in time, satisfying $t_\text{off}-t_\text{on}=\frac{mL}{2}(\frac{1}{v^{\rightarrow}}+\frac{1}{v^{\leftarrow}})$, where $m$ is any odd number, and $v^{\rightleftarrows}$ are the phase velocities in the NBM. Similarly, switching ON can take place at any moment that satisfies $t_\text{on}-t_\text{off}=\frac{mL}{v'}$, where $v'$ is the phase velocity in the dielectric medium. The only limitation on the modulation frequency are introduced by material losses, where the modulation has to be fast enough to induce field amplification that overcomes the losses to avoid overall attenuation of the fields. Relaxing the limitations on  modulation frequency and the switching speed is a significant step forward, easing the practical challenges of realizing novel phenomena based on cascaded spatiotemporal modulations.
\begin{figure}[t!]
\centerline
{\includegraphics[width=1\linewidth]{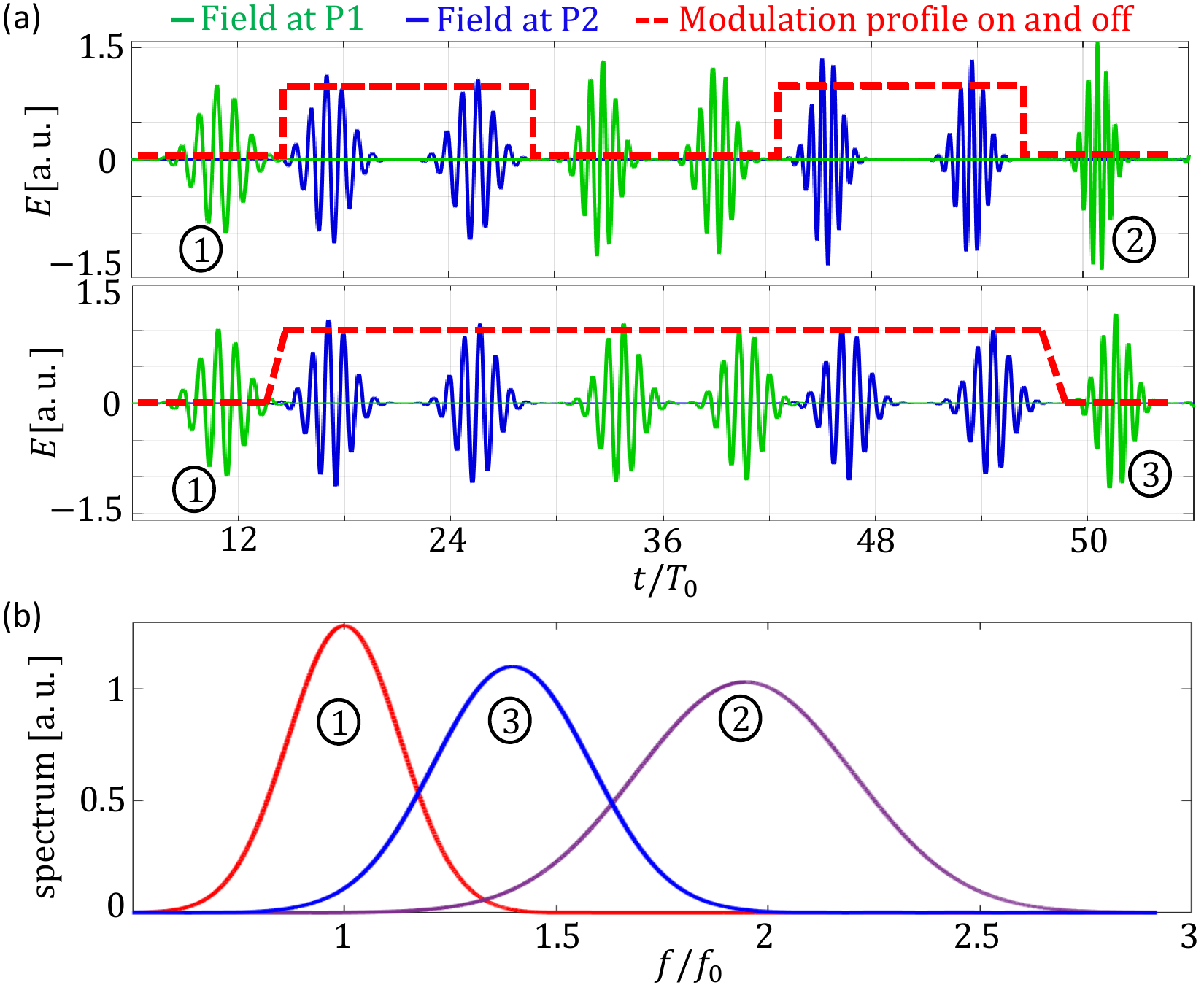}}
\caption{ (a)-(b) Temporal evolution of the electric field inside the cavity while the NBM is switched on and off. (c) Spectrum
of the initial and final pulses.} 
\label{fig_3}
\end{figure} 

Figure~\ref{fig_3} demonstrates simulations of cascaded frequency and energy amplification for a Gaussian pulse with center frequency $\omega_0$. Figure~\ref{fig_3}(a) shows the fastest possible modulation frequency with $m=1$, while Fig.~\ref{fig_3}(b) shows slower modulation with $m=2$. In addition, the switching speed for each interface in Fig.~\ref{fig_3}(b) is adiabatic. Figure~\ref{fig_3}(c) shows the spectrum of the initial and final pulses, manifesting clear frequency up-conversion that confirm the theoretical calculations. Recall that the energy gain is exactly equal to the frequency translation ratio, thus, Fig.~\ref{fig_3}(c) also illustrates the amplification coefficient. Figure~\ref{fig_3} illustrates three important points: first, cascaded amplification can be achieved with a modulation frequency much lower than the frequency of the wave $\omega_0$. Second, the switching speed (of a single interface) can be adiabatic without affecting the amplification, given that it is faster than half round trip. Finally, the amplification is inherently broadband, independent of the value of $\omega_0$.


\noindent{\textit{Conclusions}}---
This work represents a step forward towards the realization of broadband light amplification through adiabatic spatiotemporal modulation. We showed that adiabatic modulation suppresses photon-pair generation and temporal reflections, while enabling substantial gain over just a few modulation cycles. This approach holds promise for generating ultrashort, high-energy pulses and advancing the capabilities of ultrafast photonics.


\begin{acknowledgments}
M.S.M. wishes to acknowledge the support of the Research Council of Finland (Grant No.~336119). E.G., S.Y., and A.A. were supported by the Simons Foundation and the Department of Defense.
\end{acknowledgments}


\noindent{\textit{Contributions}}---
M.H.M. conceived the idea of broadband amplification through adiabatic spatiotemporal modulation. All authors contributed in developing the idea further. M.H.M. performed the theoretical calculations and numerical simulations. S.A.T. and A.A. supervised the work. All the authors contributed to the discussions of the results and the manuscript preparation.



\nocite{*}

\bibliography{ref}

\begin{thebibliography}{54}%
\makeatletter
\providecommand \@ifxundefined [1]{%
 \@ifx{#1\undefined}
}%
\providecommand \@ifnum [1]{%
 \ifnum #1\expandafter \@firstoftwo
 \else \expandafter \@secondoftwo
 \fi
}%
\providecommand \@ifx [1]{%
 \ifx #1\expandafter \@firstoftwo
 \else \expandafter \@secondoftwo
 \fi
}%
\providecommand \natexlab [1]{#1}%
\providecommand \enquote  [1]{``#1''}%
\providecommand \bibnamefont  [1]{#1}%
\providecommand \bibfnamefont [1]{#1}%
\providecommand \citenamefont [1]{#1}%
\providecommand \href@noop [0]{\@secondoftwo}%
\providecommand \href [0]{\begingroup \@sanitize@url \@href}%
\providecommand \@href[1]{\@@startlink{#1}\@@href}%
\providecommand \@@href[1]{\endgroup#1\@@endlink}%
\providecommand \@sanitize@url [0]{\catcode `\\12\catcode `\$12\catcode `\&12\catcode `\#12\catcode `\^12\catcode `\_12\catcode `\%12\relax}%
\providecommand \@@startlink[1]{}%
\providecommand \@@endlink[0]{}%
\providecommand \url  [0]{\begingroup\@sanitize@url \@url }%
\providecommand \@url [1]{\endgroup\@href {#1}{\urlprefix }}%
\providecommand \urlprefix  [0]{URL }%
\providecommand \Eprint [0]{\href }%
\providecommand \doibase [0]{https://doi.org/}%
\providecommand \selectlanguage [0]{\@gobble}%
\providecommand \bibinfo  [0]{\@secondoftwo}%
\providecommand \bibfield  [0]{\@secondoftwo}%
\providecommand \translation [1]{[#1]}%
\providecommand \BibitemOpen [0]{}%
\providecommand \bibitemStop [0]{}%
\providecommand \bibitemNoStop [0]{.\EOS\space}%
\providecommand \EOS [0]{\spacefactor3000\relax}%
\providecommand \BibitemShut  [1]{\csname bibitem#1\endcsname}%
\let\auto@bib@innerbib\@empty
\bibitem [{\citenamefont {Morgenthaler}(1958)}]{morgenthaler1958velocity}%
  \BibitemOpen
  \bibfield  {author} {\bibinfo {author} {\bibfnamefont {F.~R.}\ \bibnamefont {Morgenthaler}},\ }\bibfield  {title} {\bibinfo {title} {Velocity modulation of electromagnetic waves},\ }\href@noop {} {\bibfield  {journal} {\bibinfo  {journal} {IRE Transactions on Microwave Theory and Techniques}\ }\textbf {\bibinfo {volume} {6}},\ \bibinfo {pages} {167} (\bibinfo {year} {1958})}\BibitemShut {NoStop}%
\bibitem [{\citenamefont {Lyubarov}\ \emph {et~al.}(2022)\citenamefont {Lyubarov}, \citenamefont {Lumer}, \citenamefont {Dikopoltsev}, \citenamefont {Lustig}, \citenamefont {Sharabi},\ and\ \citenamefont {Segev}}]{lyubarov2022amplified}%
  \BibitemOpen
  \bibfield  {author} {\bibinfo {author} {\bibfnamefont {M.}~\bibnamefont {Lyubarov}}, \bibinfo {author} {\bibfnamefont {Y.}~\bibnamefont {Lumer}}, \bibinfo {author} {\bibfnamefont {A.}~\bibnamefont {Dikopoltsev}}, \bibinfo {author} {\bibfnamefont {E.}~\bibnamefont {Lustig}}, \bibinfo {author} {\bibfnamefont {Y.}~\bibnamefont {Sharabi}},\ and\ \bibinfo {author} {\bibfnamefont {M.}~\bibnamefont {Segev}},\ }\bibfield  {title} {\bibinfo {title} {Amplified emission and lasing in photonic time crystals},\ }\href {https://doi.org/10.1126/science.abo3324} {\bibfield  {journal} {\bibinfo  {journal} {Science}\ }\textbf {\bibinfo {volume} {377}},\ \bibinfo {pages} {425} (\bibinfo {year} {2022})}\BibitemShut {NoStop}%
\bibitem [{\citenamefont {Mendon\ifmmode~\mbox{\c{c}}\else \c{c}\fi{}a}\ \emph {et~al.}(2000)\citenamefont {Mendon\ifmmode~\mbox{\c{c}}\else \c{c}\fi{}a}, \citenamefont {Guerreiro},\ and\ \citenamefont {Martins}}]{PhysRevA.62.033805}%
  \BibitemOpen
  \bibfield  {author} {\bibinfo {author} {\bibfnamefont {J.~T.}\ \bibnamefont {Mendon\ifmmode~\mbox{\c{c}}\else \c{c}\fi{}a}}, \bibinfo {author} {\bibfnamefont {A.}~\bibnamefont {Guerreiro}},\ and\ \bibinfo {author} {\bibfnamefont {A.~M.}\ \bibnamefont {Martins}},\ }\bibfield  {title} {\bibinfo {title} {Quantum theory of time refraction},\ }\href {https://doi.org/10.1103/PhysRevA.62.033805} {\bibfield  {journal} {\bibinfo  {journal} {Phys. Rev. A}\ }\textbf {\bibinfo {volume} {62}},\ \bibinfo {pages} {033805} (\bibinfo {year} {2000})}\BibitemShut {NoStop}%
\bibitem [{\citenamefont {Mendon{\c{c}}a}\ and\ \citenamefont {Shukla}(2002)}]{mendoncca2002time}%
  \BibitemOpen
  \bibfield  {author} {\bibinfo {author} {\bibfnamefont {J.}~\bibnamefont {Mendon{\c{c}}a}}\ and\ \bibinfo {author} {\bibfnamefont {P.}~\bibnamefont {Shukla}},\ }\bibfield  {title} {\bibinfo {title} {Time refraction and time reflection: two basic concepts},\ }\href@noop {} {\bibfield  {journal} {\bibinfo  {journal} {Physica Scripta}\ }\textbf {\bibinfo {volume} {65}},\ \bibinfo {pages} {160} (\bibinfo {year} {2002})}\BibitemShut {NoStop}%
\bibitem [{\citenamefont {Xiao}\ \emph {et~al.}(2014{\natexlab{a}})\citenamefont {Xiao}, \citenamefont {Maywar},\ and\ \citenamefont {Agrawal}}]{Agrawal2014RTC}%
  \BibitemOpen
  \bibfield  {author} {\bibinfo {author} {\bibfnamefont {Y.}~\bibnamefont {Xiao}}, \bibinfo {author} {\bibfnamefont {D.~N.}\ \bibnamefont {Maywar}},\ and\ \bibinfo {author} {\bibfnamefont {G.~P.}\ \bibnamefont {Agrawal}},\ }\bibfield  {title} {\bibinfo {title} {Reflection and transmission of electromagnetic waves at a temporal boundary},\ }\href@noop {} {\bibfield  {journal} {\bibinfo  {journal} {Optics letters}\ }\textbf {\bibinfo {volume} {39}},\ \bibinfo {pages} {574} (\bibinfo {year} {2014}{\natexlab{a}})}\BibitemShut {NoStop}%
\bibitem [{\citenamefont {Shvartsburg}(2005)}]{Aleksandr2005}%
  \BibitemOpen
  \bibfield  {author} {\bibinfo {author} {\bibfnamefont {A.~B.}\ \bibnamefont {Shvartsburg}},\ }\bibfield  {title} {\bibinfo {title} {Optics of nonstationary media},\ }\href {https://doi.org/10.1070/PU2005v048n08ABEH002119} {\bibfield  {journal} {\bibinfo  {journal} {Physics-Uspekhi}\ }\textbf {\bibinfo {volume} {48}},\ \bibinfo {pages} {797} (\bibinfo {year} {2005})}\BibitemShut {NoStop}%
\bibitem [{\citenamefont {Mirmoosa}\ \emph {et~al.}(2025)\citenamefont {Mirmoosa}, \citenamefont {Set\"al\"a},\ and\ \citenamefont {Norrman}}]{sajjad:Q}%
  \BibitemOpen
  \bibfield  {author} {\bibinfo {author} {\bibfnamefont {M.~S.}\ \bibnamefont {Mirmoosa}}, \bibinfo {author} {\bibfnamefont {T.}~\bibnamefont {Set\"al\"a}},\ and\ \bibinfo {author} {\bibfnamefont {A.}~\bibnamefont {Norrman}},\ }\bibfield  {title} {\bibinfo {title} {Quantum state engineering and photon statistics at electromagnetic time interfaces},\ }\href {https://doi.org/10.1103/PhysRevResearch.7.013120} {\bibfield  {journal} {\bibinfo  {journal} {Phys. Rev. Res.}\ }\textbf {\bibinfo {volume} {7}},\ \bibinfo {pages} {013120} (\bibinfo {year} {2025})}\BibitemShut {NoStop}%
\bibitem [{\citenamefont {Moussa}\ \emph {et~al.}(2023{\natexlab{a}})\citenamefont {Moussa}, \citenamefont {Xu}, \citenamefont {Yin}, \citenamefont {Galiffi}, \citenamefont {Ra’di},\ and\ \citenamefont {Alù}}]{Alu_TL}%
  \BibitemOpen
  \bibfield  {author} {\bibinfo {author} {\bibfnamefont {H.}~\bibnamefont {Moussa}}, \bibinfo {author} {\bibfnamefont {G.}~\bibnamefont {Xu}}, \bibinfo {author} {\bibfnamefont {S.}~\bibnamefont {Yin}}, \bibinfo {author} {\bibfnamefont {E.}~\bibnamefont {Galiffi}}, \bibinfo {author} {\bibfnamefont {Y.}~\bibnamefont {Ra’di}},\ and\ \bibinfo {author} {\bibfnamefont {A.}~\bibnamefont {Alù}},\ }\bibfield  {title} {\bibinfo {title} {Observation of temporal reflection and broadband frequency translation at photonic time interfaces},\ }\href {https://doi.org/10.1038/s41567-023-01975-y} {\bibfield  {journal} {\bibinfo  {journal} {Nature Physics}\ }\textbf {\bibinfo {volume} {19}},\ \bibinfo {pages} {863} (\bibinfo {year} {2023}{\natexlab{a}})}\BibitemShut {NoStop}%
\bibitem [{\citenamefont {Jones}\ \emph {et~al.}(2024{\natexlab{a}})\citenamefont {Jones}, \citenamefont {Kildishev}, \citenamefont {Segev},\ and\ \citenamefont {Peroulis}}]{jones2024time}%
  \BibitemOpen
  \bibfield  {author} {\bibinfo {author} {\bibfnamefont {T.~R.}\ \bibnamefont {Jones}}, \bibinfo {author} {\bibfnamefont {A.~V.}\ \bibnamefont {Kildishev}}, \bibinfo {author} {\bibfnamefont {M.}~\bibnamefont {Segev}},\ and\ \bibinfo {author} {\bibfnamefont {D.}~\bibnamefont {Peroulis}},\ }\bibfield  {title} {\bibinfo {title} {Time-reflection of microwaves by a fast optically-controlled time-boundary},\ }\href@noop {} {\bibfield  {journal} {\bibinfo  {journal} {Nature Communications}\ }\textbf {\bibinfo {volume} {15}},\ \bibinfo {pages} {6786} (\bibinfo {year} {2024}{\natexlab{a}})}\BibitemShut {NoStop}%
\bibitem [{\citenamefont {Xiao}\ \emph {et~al.}(2014{\natexlab{b}})\citenamefont {Xiao}, \citenamefont {Maywar},\ and\ \citenamefont {Agrawal}}]{Xiao:14}%
  \BibitemOpen
  \bibfield  {author} {\bibinfo {author} {\bibfnamefont {Y.}~\bibnamefont {Xiao}}, \bibinfo {author} {\bibfnamefont {D.~N.}\ \bibnamefont {Maywar}},\ and\ \bibinfo {author} {\bibfnamefont {G.~P.}\ \bibnamefont {Agrawal}},\ }\bibfield  {title} {\bibinfo {title} {Reflection and transmission of electromagnetic waves at a temporal boundary},\ }\href {https://doi.org/10.1364/OL.39.000574} {\bibfield  {journal} {\bibinfo  {journal} {Opt. Lett.}\ }\textbf {\bibinfo {volume} {39}},\ \bibinfo {pages} {574} (\bibinfo {year} {2014}{\natexlab{b}})}\BibitemShut {NoStop}%
\bibitem [{\citenamefont {Mostafa}\ \emph {et~al.}(2024)\citenamefont {Mostafa}, \citenamefont {Mirmoosa}, \citenamefont {Sidorenko}, \citenamefont {Asadchy},\ and\ \citenamefont {Tretyakov}}]{Mostafa:24}%
  \BibitemOpen
  \bibfield  {author} {\bibinfo {author} {\bibfnamefont {M.~H.}\ \bibnamefont {Mostafa}}, \bibinfo {author} {\bibfnamefont {M.~S.}\ \bibnamefont {Mirmoosa}}, \bibinfo {author} {\bibfnamefont {M.~S.}\ \bibnamefont {Sidorenko}}, \bibinfo {author} {\bibfnamefont {V.~S.}\ \bibnamefont {Asadchy}},\ and\ \bibinfo {author} {\bibfnamefont {S.~A.}\ \bibnamefont {Tretyakov}},\ }\bibfield  {title} {\bibinfo {title} {Temporal interfaces in complex electromagnetic materials: an overview [invited]},\ }\href {https://doi.org/10.1364/OME.516179} {\bibfield  {journal} {\bibinfo  {journal} {Opt. Mater. Express}\ }\textbf {\bibinfo {volume} {14}},\ \bibinfo {pages} {1103} (\bibinfo {year} {2024})}\BibitemShut {NoStop}%
\bibitem [{\citenamefont {Ramaccia}\ \emph {et~al.}(2020)\citenamefont {Ramaccia}, \citenamefont {Toscano},\ and\ \citenamefont {Bilotti}}]{ramaccia2020light}%
  \BibitemOpen
  \bibfield  {author} {\bibinfo {author} {\bibfnamefont {D.}~\bibnamefont {Ramaccia}}, \bibinfo {author} {\bibfnamefont {A.}~\bibnamefont {Toscano}},\ and\ \bibinfo {author} {\bibfnamefont {F.}~\bibnamefont {Bilotti}},\ }\bibfield  {title} {\bibinfo {title} {Light propagation through metamaterial temporal slabs: reflection, refraction, and special cases},\ }\href@noop {} {\bibfield  {journal} {\bibinfo  {journal} {Optics Letters}\ }\textbf {\bibinfo {volume} {45}},\ \bibinfo {pages} {5836} (\bibinfo {year} {2020})}\BibitemShut {NoStop}%
\bibitem [{\citenamefont {Galiffi}\ \emph {et~al.}(2019)\citenamefont {Galiffi}, \citenamefont {Huidobro},\ and\ \citenamefont {Pendry}}]{Emanuele_luminal}%
  \BibitemOpen
  \bibfield  {author} {\bibinfo {author} {\bibfnamefont {E.}~\bibnamefont {Galiffi}}, \bibinfo {author} {\bibfnamefont {P.~A.}\ \bibnamefont {Huidobro}},\ and\ \bibinfo {author} {\bibfnamefont {J.~B.}\ \bibnamefont {Pendry}},\ }\bibfield  {title} {\bibinfo {title} {Broadband nonreciprocal amplification in luminal metamaterials},\ }\href {https://doi.org/10.1103/PhysRevLett.123.206101} {\bibfield  {journal} {\bibinfo  {journal} {Phys. Rev. Lett.}\ }\textbf {\bibinfo {volume} {123}},\ \bibinfo {pages} {206101} (\bibinfo {year} {2019})}\BibitemShut {NoStop}%
\bibitem [{\citenamefont {Ptitcyn}\ \emph {et~al.}(2023)\citenamefont {Ptitcyn}, \citenamefont {Mirmoosa}, \citenamefont {Sotoodehfar},\ and\ \citenamefont {Tretyakov}}]{Ptitcyn2023Tutorial}%
  \BibitemOpen
  \bibfield  {author} {\bibinfo {author} {\bibfnamefont {G.}~\bibnamefont {Ptitcyn}}, \bibinfo {author} {\bibfnamefont {M.~S.}\ \bibnamefont {Mirmoosa}}, \bibinfo {author} {\bibfnamefont {A.}~\bibnamefont {Sotoodehfar}},\ and\ \bibinfo {author} {\bibfnamefont {S.~A.}\ \bibnamefont {Tretyakov}},\ }\bibfield  {title} {\bibinfo {title} {A tutorial on the basics of time-varying electromagnetic systems and circuits: Historic overview and basic concepts of time-modulation},\ }\href@noop {} {\bibfield  {journal} {\bibinfo  {journal} {IEEE Antennas and Propagation Magazine}\ }\textbf {\bibinfo {volume} {65}},\ \bibinfo {pages} {10} (\bibinfo {year} {2023})}\BibitemShut {NoStop}%
\bibitem [{\citenamefont {Wang}\ \emph {et~al.}(2023)\citenamefont {Wang}, \citenamefont {Mirmoosa}, \citenamefont {Asadchy}, \citenamefont {Rockstuhl}, \citenamefont {Fan},\ and\ \citenamefont {Tretyakov}}]{WaMIR23PTC}%
  \BibitemOpen
  \bibfield  {author} {\bibinfo {author} {\bibfnamefont {X.}~\bibnamefont {Wang}}, \bibinfo {author} {\bibfnamefont {M.~S.}\ \bibnamefont {Mirmoosa}}, \bibinfo {author} {\bibfnamefont {V.~S.}\ \bibnamefont {Asadchy}}, \bibinfo {author} {\bibfnamefont {C.}~\bibnamefont {Rockstuhl}}, \bibinfo {author} {\bibfnamefont {S.}~\bibnamefont {Fan}},\ and\ \bibinfo {author} {\bibfnamefont {S.~A.}\ \bibnamefont {Tretyakov}},\ }\bibfield  {title} {\bibinfo {title} {Metasurface-based realization of photonic time crystals},\ }\href@noop {} {\bibfield  {journal} {\bibinfo  {journal} {Science Advances}\ }\textbf {\bibinfo {volume} {9}},\ \bibinfo {pages} {eadg7541} (\bibinfo {year} {2023})}\BibitemShut {NoStop}%
\bibitem [{\citenamefont {Zurita-S\'anchez}\ \emph {et~al.}(2009)\citenamefont {Zurita-S\'anchez}, \citenamefont {Halevi},\ and\ \citenamefont {Cervantes-Gonz\'alez}}]{Halevi9PTC}%
  \BibitemOpen
  \bibfield  {author} {\bibinfo {author} {\bibfnamefont {J.~R.}\ \bibnamefont {Zurita-S\'anchez}}, \bibinfo {author} {\bibfnamefont {P.}~\bibnamefont {Halevi}},\ and\ \bibinfo {author} {\bibfnamefont {J.~C.}\ \bibnamefont {Cervantes-Gonz\'alez}},\ }\bibfield  {title} {\bibinfo {title} {Reflection and transmission of a wave incident on a slab with a time-periodic dielectric function $\epsilon(t)$},\ }\href {https://doi.org/10.1103/PhysRevA.79.053821} {\bibfield  {journal} {\bibinfo  {journal} {Phys. Rev. A}\ }\textbf {\bibinfo {volume} {79}},\ \bibinfo {pages} {053821} (\bibinfo {year} {2009})}\BibitemShut {NoStop}%
\bibitem [{\citenamefont {Lustig}\ \emph {et~al.}(2018)\citenamefont {Lustig}, \citenamefont {Sharabi},\ and\ \citenamefont {Segev}}]{Segev8PTC}%
  \BibitemOpen
  \bibfield  {author} {\bibinfo {author} {\bibfnamefont {E.}~\bibnamefont {Lustig}}, \bibinfo {author} {\bibfnamefont {Y.}~\bibnamefont {Sharabi}},\ and\ \bibinfo {author} {\bibfnamefont {M.}~\bibnamefont {Segev}},\ }\bibfield  {title} {\bibinfo {title} {Topological aspects of photonic time crystals},\ }\href@noop {} {\bibfield  {journal} {\bibinfo  {journal} {Optica}\ }\textbf {\bibinfo {volume} {5}},\ \bibinfo {pages} {1390} (\bibinfo {year} {2018})}\BibitemShut {NoStop}%
\bibitem [{\citenamefont {Boltasseva}\ \emph {et~al.}(2024)\citenamefont {Boltasseva}, \citenamefont {Shalaev},\ and\ \citenamefont {Segev}}]{Boltasseva:24}%
  \BibitemOpen
  \bibfield  {author} {\bibinfo {author} {\bibfnamefont {A.}~\bibnamefont {Boltasseva}}, \bibinfo {author} {\bibfnamefont {V.~M.}\ \bibnamefont {Shalaev}},\ and\ \bibinfo {author} {\bibfnamefont {M.}~\bibnamefont {Segev}},\ }\bibfield  {title} {\bibinfo {title} {Photonic time crystals: from fundamental insights to novel applications: opinion},\ }\href {https://doi.org/10.1364/OME.511801} {\bibfield  {journal} {\bibinfo  {journal} {Opt. Mater. Express}\ }\textbf {\bibinfo {volume} {14}},\ \bibinfo {pages} {592} (\bibinfo {year} {2024})}\BibitemShut {NoStop}%
\bibitem [{\citenamefont {Asgari}\ \emph {et~al.}(2024)\citenamefont {Asgari}, \citenamefont {Garg}, \citenamefont {Wang}, \citenamefont {Mirmoosa}, \citenamefont {Rockstuhl},\ and\ \citenamefont {Asadchy}}]{asgari2024photonic}%
  \BibitemOpen
  \bibfield  {author} {\bibinfo {author} {\bibfnamefont {M.~M.}\ \bibnamefont {Asgari}}, \bibinfo {author} {\bibfnamefont {P.}~\bibnamefont {Garg}}, \bibinfo {author} {\bibfnamefont {X.}~\bibnamefont {Wang}}, \bibinfo {author} {\bibfnamefont {M.~S.}\ \bibnamefont {Mirmoosa}}, \bibinfo {author} {\bibfnamefont {C.}~\bibnamefont {Rockstuhl}},\ and\ \bibinfo {author} {\bibfnamefont {V.}~\bibnamefont {Asadchy}},\ }\bibfield  {title} {\bibinfo {title} {Theory and applications of photonic time crystals: a tutorial},\ }\href {https://doi.org/10.1364/AOP.525163} {\bibfield  {journal} {\bibinfo  {journal} {Adv. Opt. Photon.}\ }\textbf {\bibinfo {volume} {16}},\ \bibinfo {pages} {958} (\bibinfo {year} {2024})}\BibitemShut {NoStop}%
\bibitem [{\citenamefont {Moussa}\ \emph {et~al.}(2023{\natexlab{b}})\citenamefont {Moussa}, \citenamefont {Xu}, \citenamefont {Yin} \emph {et~al.}}]{moussa2023}%
  \BibitemOpen
  \bibfield  {author} {\bibinfo {author} {\bibfnamefont {H.}~\bibnamefont {Moussa}}, \bibinfo {author} {\bibfnamefont {G.}~\bibnamefont {Xu}}, \bibinfo {author} {\bibfnamefont {S.}~\bibnamefont {Yin}}, \emph {et~al.},\ }\bibfield  {title} {\bibinfo {title} {Observation of temporal reflection and broadband frequency translation at photonic time interfaces},\ }\href {https://doi.org/10.1038/s41567-023-02021-6} {\bibfield  {journal} {\bibinfo  {journal} {Nature Physics}\ }\textbf {\bibinfo {volume} {19}},\ \bibinfo {pages} {863} (\bibinfo {year} {2023}{\natexlab{b}})}\BibitemShut {NoStop}%
\bibitem [{\citenamefont {Jones}\ \emph {et~al.}(2024{\natexlab{b}})\citenamefont {Jones}, \citenamefont {Kildishev}, \citenamefont {Segev} \emph {et~al.}}]{jones2024}%
  \BibitemOpen
  \bibfield  {author} {\bibinfo {author} {\bibfnamefont {T.}~\bibnamefont {Jones}}, \bibinfo {author} {\bibfnamefont {A.}~\bibnamefont {Kildishev}}, \bibinfo {author} {\bibfnamefont {M.}~\bibnamefont {Segev}}, \emph {et~al.},\ }\bibfield  {title} {\bibinfo {title} {Time-reflection of microwaves by a fast optically-controlled time-boundary},\ }\href {https://doi.org/10.1038/s41467-024-06886-3} {\bibfield  {journal} {\bibinfo  {journal} {Nature Communications}\ }\textbf {\bibinfo {volume} {15}},\ \bibinfo {pages} {6786} (\bibinfo {year} {2024}{\natexlab{b}})}\BibitemShut {NoStop}%
\bibitem [{\citenamefont {Galiffi}\ \emph {et~al.}(2022{\natexlab{a}})\citenamefont {Galiffi}, \citenamefont {Yin},\ and\ \citenamefont {Alú}}]{tapered}%
  \BibitemOpen
  \bibfield  {author} {\bibinfo {author} {\bibfnamefont {E.}~\bibnamefont {Galiffi}}, \bibinfo {author} {\bibfnamefont {S.}~\bibnamefont {Yin}},\ and\ \bibinfo {author} {\bibfnamefont {A.}~\bibnamefont {Alú}},\ }\bibfield  {title} {\bibinfo {title} {Tapered photonic switching},\ }\href {https://doi.org/doi:10.1515/nanoph-2022-0200} {\bibfield  {journal} {\bibinfo  {journal} {Nanophotonics}\ }\textbf {\bibinfo {volume} {11}},\ \bibinfo {pages} {3575} (\bibinfo {year} {2022}{\natexlab{a}})}\BibitemShut {NoStop}%
\bibitem [{\citenamefont {Hayran}\ \emph {et~al.}(2022)\citenamefont {Hayran}, \citenamefont {Khurgin},\ and\ \citenamefont {Monticone}}]{Hayran:22}%
  \BibitemOpen
  \bibfield  {author} {\bibinfo {author} {\bibfnamefont {Z.}~\bibnamefont {Hayran}}, \bibinfo {author} {\bibfnamefont {J.~B.}\ \bibnamefont {Khurgin}},\ and\ \bibinfo {author} {\bibfnamefont {F.}~\bibnamefont {Monticone}},\ }\bibfield  {title} {\bibinfo {title} {$ \hbar \omega$ versus $\hbar k$: dispersion and energy constraints on time-varying photonic materials and time crystals [invited]},\ }\href {https://opg.optica.org/ome/abstract.cfm?URI=ome-12-10-3904} {\bibfield  {journal} {\bibinfo  {journal} {Opt. Mater. Express}\ }\textbf {\bibinfo {volume} {12}},\ \bibinfo {pages} {3904} (\bibinfo {year} {2022})}\BibitemShut {NoStop}%
\bibitem [{\citenamefont {Engheta}(2023)}]{engheta2023four}%
  \BibitemOpen
  \bibfield  {author} {\bibinfo {author} {\bibfnamefont {N.}~\bibnamefont {Engheta}},\ }\bibfield  {title} {\bibinfo {title} {Four-dimensional optics using time-varying metamaterials},\ }\href@noop {} {\bibfield  {journal} {\bibinfo  {journal} {Science}\ }\textbf {\bibinfo {volume} {379}},\ \bibinfo {pages} {1190} (\bibinfo {year} {2023})}\BibitemShut {NoStop}%
\bibitem [{\citenamefont {Galiffi}\ \emph {et~al.}(2022{\natexlab{b}})\citenamefont {Galiffi}, \citenamefont {Tirole}, \citenamefont {Yin}, \citenamefont {Li}, \citenamefont {Vezzoli}, \citenamefont {Huidobro}, \citenamefont {Silveirinha}, \citenamefont {Sapienza}, \citenamefont {Al{\`u}},\ and\ \citenamefont {Pendry}}]{galiffi2022photonics}%
  \BibitemOpen
  \bibfield  {author} {\bibinfo {author} {\bibfnamefont {E.}~\bibnamefont {Galiffi}}, \bibinfo {author} {\bibfnamefont {R.}~\bibnamefont {Tirole}}, \bibinfo {author} {\bibfnamefont {S.}~\bibnamefont {Yin}}, \bibinfo {author} {\bibfnamefont {H.}~\bibnamefont {Li}}, \bibinfo {author} {\bibfnamefont {S.}~\bibnamefont {Vezzoli}}, \bibinfo {author} {\bibfnamefont {P.~A.}\ \bibnamefont {Huidobro}}, \bibinfo {author} {\bibfnamefont {M.~G.}\ \bibnamefont {Silveirinha}}, \bibinfo {author} {\bibfnamefont {R.}~\bibnamefont {Sapienza}}, \bibinfo {author} {\bibfnamefont {A.}~\bibnamefont {Al{\`u}}},\ and\ \bibinfo {author} {\bibfnamefont {J.}~\bibnamefont {Pendry}},\ }\bibfield  {title} {\bibinfo {title} {Photonics of time-varying media},\ }\href@noop {} {\bibfield  {journal} {\bibinfo  {journal} {Advanced Photonics}\ }\textbf {\bibinfo {volume} {4}},\ \bibinfo {pages} {014002} (\bibinfo {year} {2022}{\natexlab{b}})}\BibitemShut {NoStop}%
\bibitem [{\citenamefont {Caloz}\ and\ \citenamefont {Deck-Léger}(2020{\natexlab{a}})}]{Caloz:1}%
  \BibitemOpen
  \bibfield  {author} {\bibinfo {author} {\bibfnamefont {C.}~\bibnamefont {Caloz}}\ and\ \bibinfo {author} {\bibfnamefont {Z.-L.}\ \bibnamefont {Deck-Léger}},\ }\bibfield  {title} {\bibinfo {title} {Spacetime metamaterials—part i: General concepts},\ }\href@noop {} {\bibfield  {journal} {\bibinfo  {journal} {IEEE Transactions on Antennas and Propagation}\ }\textbf {\bibinfo {volume} {68}},\ \bibinfo {pages} {1569} (\bibinfo {year} {2020}{\natexlab{a}})}\BibitemShut {NoStop}%
\bibitem [{\citenamefont {Caloz}\ and\ \citenamefont {Deck-Léger}(2020{\natexlab{b}})}]{Caloz:2}%
  \BibitemOpen
  \bibfield  {author} {\bibinfo {author} {\bibfnamefont {C.}~\bibnamefont {Caloz}}\ and\ \bibinfo {author} {\bibfnamefont {Z.-L.}\ \bibnamefont {Deck-Léger}},\ }\bibfield  {title} {\bibinfo {title} {Spacetime metamaterials—part ii: Theory and applications},\ }\href@noop {} {\bibfield  {journal} {\bibinfo  {journal} {IEEE Transactions on Antennas and Propagation}\ }\textbf {\bibinfo {volume} {68}},\ \bibinfo {pages} {1583} (\bibinfo {year} {2020}{\natexlab{b}})}\BibitemShut {NoStop}%
\bibitem [{\citenamefont {Guo}\ \emph {et~al.}(2019)\citenamefont {Guo}, \citenamefont {Ding}, \citenamefont {Duan} \emph {et~al.}}]{guo2019nonreciprocal}%
  \BibitemOpen
  \bibfield  {author} {\bibinfo {author} {\bibfnamefont {X.}~\bibnamefont {Guo}}, \bibinfo {author} {\bibfnamefont {Y.}~\bibnamefont {Ding}}, \bibinfo {author} {\bibfnamefont {Y.}~\bibnamefont {Duan}}, \emph {et~al.},\ }\bibfield  {title} {\bibinfo {title} {Nonreciprocal metasurface with space--time phase modulation},\ }\href {https://doi.org/10.1038/s41377-019-0225-z} {\bibfield  {journal} {\bibinfo  {journal} {Light: Science \& Applications}\ }\textbf {\bibinfo {volume} {8}},\ \bibinfo {pages} {123} (\bibinfo {year} {2019})}\BibitemShut {NoStop}%
\bibitem [{\citenamefont {Harwood}\ \emph {et~al.}(2024)\citenamefont {Harwood}, \citenamefont {Vezzoli}, \citenamefont {Raziman}, \citenamefont {Hooper}, \citenamefont {Tirole}, \citenamefont {Wu}, \citenamefont {Maier}, \citenamefont {Pendry}, \citenamefont {Horsley},\ and\ \citenamefont {Sapienza}}]{harwood2024super}%
  \BibitemOpen
  \bibfield  {author} {\bibinfo {author} {\bibfnamefont {A.~C.}\ \bibnamefont {Harwood}}, \bibinfo {author} {\bibfnamefont {S.}~\bibnamefont {Vezzoli}}, \bibinfo {author} {\bibfnamefont {T.~V.}\ \bibnamefont {Raziman}}, \bibinfo {author} {\bibfnamefont {C.}~\bibnamefont {Hooper}}, \bibinfo {author} {\bibfnamefont {R.}~\bibnamefont {Tirole}}, \bibinfo {author} {\bibfnamefont {F.}~\bibnamefont {Wu}}, \bibinfo {author} {\bibfnamefont {S.~A.}\ \bibnamefont {Maier}}, \bibinfo {author} {\bibfnamefont {J.~B.}\ \bibnamefont {Pendry}}, \bibinfo {author} {\bibfnamefont {S.~A.~R.}\ \bibnamefont {Horsley}},\ and\ \bibinfo {author} {\bibfnamefont {R.}~\bibnamefont {Sapienza}},\ }\bibfield  {title} {\bibinfo {title} {Super-luminal synthetic motion with a space-time optical metasurface},\ }\href@noop {} {\bibfield  {journal} {\bibinfo  {journal} {arXiv:2407.10809}\ } (\bibinfo {year} {2024})}\BibitemShut {NoStop}%
\bibitem [{\citenamefont {Zhou}\ \emph {et~al.}(2020)\citenamefont {Zhou}, \citenamefont {Alam}, \citenamefont {Karimi}, \citenamefont {Upham}, \citenamefont {Reshef}, \citenamefont {Liu}, \citenamefont {Willner},\ and\ \citenamefont {Boyd}}]{zhou2020broadband}%
  \BibitemOpen
  \bibfield  {author} {\bibinfo {author} {\bibfnamefont {Y.}~\bibnamefont {Zhou}}, \bibinfo {author} {\bibfnamefont {M.~Z.}\ \bibnamefont {Alam}}, \bibinfo {author} {\bibfnamefont {M.}~\bibnamefont {Karimi}}, \bibinfo {author} {\bibfnamefont {J.}~\bibnamefont {Upham}}, \bibinfo {author} {\bibfnamefont {O.}~\bibnamefont {Reshef}}, \bibinfo {author} {\bibfnamefont {C.}~\bibnamefont {Liu}}, \bibinfo {author} {\bibfnamefont {A.~E.}\ \bibnamefont {Willner}},\ and\ \bibinfo {author} {\bibfnamefont {R.~W.}\ \bibnamefont {Boyd}},\ }\bibfield  {title} {\bibinfo {title} {Broadband frequency translation through time refraction in an epsilon-near-zero material},\ }\href {https://doi.org/10.1038/s41467-020-15917-3} {\bibfield  {journal} {\bibinfo  {journal} {Nature Communications}\ }\textbf {\bibinfo {volume} {11}},\ \bibinfo {pages} {2180} (\bibinfo {year} {2020})}\BibitemShut {NoStop}%
\bibitem [{\citenamefont {Lustig}\ \emph {et~al.}(2023)\citenamefont {Lustig}, \citenamefont {Segal}, \citenamefont {Saha}, \citenamefont {Bordo}, \citenamefont {Chowdhury}, \citenamefont {Sharabi}, \citenamefont {Fleischer}, \citenamefont {Boltasseva}, \citenamefont {Cohen}, \citenamefont {Shalaev},\ and\ \citenamefont {Segev}}]{segev:single}%
  \BibitemOpen
  \bibfield  {author} {\bibinfo {author} {\bibfnamefont {E.}~\bibnamefont {Lustig}}, \bibinfo {author} {\bibfnamefont {O.}~\bibnamefont {Segal}}, \bibinfo {author} {\bibfnamefont {S.}~\bibnamefont {Saha}}, \bibinfo {author} {\bibfnamefont {E.}~\bibnamefont {Bordo}}, \bibinfo {author} {\bibfnamefont {S.~N.}\ \bibnamefont {Chowdhury}}, \bibinfo {author} {\bibfnamefont {Y.}~\bibnamefont {Sharabi}}, \bibinfo {author} {\bibfnamefont {A.}~\bibnamefont {Fleischer}}, \bibinfo {author} {\bibfnamefont {A.}~\bibnamefont {Boltasseva}}, \bibinfo {author} {\bibfnamefont {O.}~\bibnamefont {Cohen}}, \bibinfo {author} {\bibfnamefont {V.~M.}\ \bibnamefont {Shalaev}},\ and\ \bibinfo {author} {\bibfnamefont {M.}~\bibnamefont {Segev}},\ }\bibfield  {title} {\bibinfo {title} {Time-refraction optics with single cycle modulation},\ }\href {https://doi.org/doi:10.1515/nanoph-2023-0126} {\bibfield  {journal} {\bibinfo  {journal} {Nanophotonics}\ }\textbf {\bibinfo {volume} {12}},\ \bibinfo {pages} {2221} (\bibinfo {year}
  {2023})}\BibitemShut {NoStop}%
\bibitem [{\citenamefont {Tirole}\ \emph {et~al.}(2023)\citenamefont {Tirole}, \citenamefont {Vezzoli}, \citenamefont {Galiffi} \emph {et~al.}}]{tirole2023double}%
  \BibitemOpen
  \bibfield  {author} {\bibinfo {author} {\bibfnamefont {R.}~\bibnamefont {Tirole}}, \bibinfo {author} {\bibfnamefont {S.}~\bibnamefont {Vezzoli}}, \bibinfo {author} {\bibfnamefont {E.}~\bibnamefont {Galiffi}}, \emph {et~al.},\ }\bibfield  {title} {\bibinfo {title} {Double-slit time diffraction at optical frequencies},\ }\href {https://doi.org/10.1038/s41567-023-01993-w} {\bibfield  {journal} {\bibinfo  {journal} {Nature Physics}\ }\textbf {\bibinfo {volume} {19}},\ \bibinfo {pages} {999} (\bibinfo {year} {2023})}\BibitemShut {NoStop}%
\bibitem [{\citenamefont {Liu}\ \emph {et~al.}(2021)\citenamefont {Liu}, \citenamefont {Alam}, \citenamefont {Pang} \emph {et~al.}}]{photon_acc}%
  \BibitemOpen
  \bibfield  {author} {\bibinfo {author} {\bibfnamefont {C.}~\bibnamefont {Liu}}, \bibinfo {author} {\bibfnamefont {M.~Z.}\ \bibnamefont {Alam}}, \bibinfo {author} {\bibfnamefont {K.}~\bibnamefont {Pang}}, \emph {et~al.},\ }\bibfield  {title} {\bibinfo {title} {Photon acceleration using a time-varying epsilon-near-zero metasurface},\ }\href@noop {} {\bibfield  {journal} {\bibinfo  {journal} {ACS Photonics}\ }\textbf {\bibinfo {volume} {8}},\ \bibinfo {pages} {716} (\bibinfo {year} {2021})}\BibitemShut {NoStop}%
\bibitem [{\citenamefont {Pang}\ \emph {et~al.}(2021)\citenamefont {Pang}, \citenamefont {Alam}, \citenamefont {Zhou}, \citenamefont {Liu} \emph {et~al.}}]{adiabatic_f}%
  \BibitemOpen
  \bibfield  {author} {\bibinfo {author} {\bibfnamefont {K.}~\bibnamefont {Pang}}, \bibinfo {author} {\bibfnamefont {M.~Z.}\ \bibnamefont {Alam}}, \bibinfo {author} {\bibfnamefont {Y.}~\bibnamefont {Zhou}}, \bibinfo {author} {\bibfnamefont {C.}~\bibnamefont {Liu}}, \emph {et~al.},\ }\bibfield  {title} {\bibinfo {title} {Adiabatic frequency conversion using a time-varying epsilon-near-zero metasurface},\ }\href@noop {} {\bibfield  {journal} {\bibinfo  {journal} {Nano Letters}\ }\textbf {\bibinfo {volume} {21}},\ \bibinfo {pages} {5907} (\bibinfo {year} {2021})}\BibitemShut {NoStop}%
\bibitem [{\citenamefont {Hadad}\ \emph {et~al.}(2015)\citenamefont {Hadad}, \citenamefont {Sounas},\ and\ \citenamefont {Alu}}]{Yakir2015}%
  \BibitemOpen
  \bibfield  {author} {\bibinfo {author} {\bibfnamefont {Y.}~\bibnamefont {Hadad}}, \bibinfo {author} {\bibfnamefont {D.~L.}\ \bibnamefont {Sounas}},\ and\ \bibinfo {author} {\bibfnamefont {A.}~\bibnamefont {Alu}},\ }\bibfield  {title} {\bibinfo {title} {Space-time gradient metasurfaces},\ }\href {https://doi.org/10.1103/PhysRevB.92.100304} {\bibfield  {journal} {\bibinfo  {journal} {Phys. Rev. B}\ }\textbf {\bibinfo {volume} {92}},\ \bibinfo {pages} {100304} (\bibinfo {year} {2015})}\BibitemShut {NoStop}%
\bibitem [{\citenamefont {Huidobro}\ \emph {et~al.}(2019)\citenamefont {Huidobro}, \citenamefont {Galiffi}, \citenamefont {Guenneau}, \citenamefont {Craster},\ and\ \citenamefont {Pendry}}]{Paloma_drag}%
  \BibitemOpen
  \bibfield  {author} {\bibinfo {author} {\bibfnamefont {P.~A.}\ \bibnamefont {Huidobro}}, \bibinfo {author} {\bibfnamefont {E.}~\bibnamefont {Galiffi}}, \bibinfo {author} {\bibfnamefont {S.}~\bibnamefont {Guenneau}}, \bibinfo {author} {\bibfnamefont {R.~V.}\ \bibnamefont {Craster}},\ and\ \bibinfo {author} {\bibfnamefont {J.~B.}\ \bibnamefont {Pendry}},\ }\bibfield  {title} {\bibinfo {title} {Fresnel drag in space–time-modulated metamaterials},\ }\href@noop {} {\bibfield  {journal} {\bibinfo  {journal} {Proceedings of the National Academy of Sciences of the United States of America}\ }\textbf {\bibinfo {volume} {116}},\ \bibinfo {pages} {24943} (\bibinfo {year} {2019})}\BibitemShut {NoStop}%
\bibitem [{\citenamefont {Pendry}\ \emph {et~al.}(2021{\natexlab{a}})\citenamefont {Pendry}, \citenamefont {Galiffi},\ and\ \citenamefont {Huidobro}}]{Pendry:21}%
  \BibitemOpen
  \bibfield  {author} {\bibinfo {author} {\bibfnamefont {J.~B.}\ \bibnamefont {Pendry}}, \bibinfo {author} {\bibfnamefont {E.}~\bibnamefont {Galiffi}},\ and\ \bibinfo {author} {\bibfnamefont {P.~A.}\ \bibnamefont {Huidobro}},\ }\bibfield  {title} {\bibinfo {title} {Gain in time-dependent media---a new mechanism},\ }\href@noop {} {\bibfield  {journal} {\bibinfo  {journal} {J. Opt. Soc. Am. B}\ }\textbf {\bibinfo {volume} {38}},\ \bibinfo {pages} {3360} (\bibinfo {year} {2021}{\natexlab{a}})}\BibitemShut {NoStop}%
\bibitem [{\citenamefont {Pendry}\ \emph {et~al.}(2021{\natexlab{b}})\citenamefont {Pendry}, \citenamefont {Galiffi},\ and\ \citenamefont {Huidobro}}]{Pendry_forceline}%
  \BibitemOpen
  \bibfield  {author} {\bibinfo {author} {\bibfnamefont {J.~B.}\ \bibnamefont {Pendry}}, \bibinfo {author} {\bibfnamefont {E.}~\bibnamefont {Galiffi}},\ and\ \bibinfo {author} {\bibfnamefont {P.~A.}\ \bibnamefont {Huidobro}},\ }\bibfield  {title} {\bibinfo {title} {Gain mechanism in time-dependent media},\ }\href {https://doi.org/10.1364/OPTICA.425582} {\bibfield  {journal} {\bibinfo  {journal} {Optica}\ }\textbf {\bibinfo {volume} {8}},\ \bibinfo {pages} {636} (\bibinfo {year} {2021}{\natexlab{b}})}\BibitemShut {NoStop}%
\bibitem [{\citenamefont {Pendry}\ \emph {et~al.}(2022)\citenamefont {Pendry}, \citenamefont {Galiffi},\ and\ \citenamefont {Huidobro}}]{Pendry:con}%
  \BibitemOpen
  \bibfield  {author} {\bibinfo {author} {\bibfnamefont {J.~B.}\ \bibnamefont {Pendry}}, \bibinfo {author} {\bibfnamefont {E.}~\bibnamefont {Galiffi}},\ and\ \bibinfo {author} {\bibfnamefont {P.~A.}\ \bibnamefont {Huidobro}},\ }\bibfield  {title} {\bibinfo {title} {Photon conservation in trans-luminal metamaterials},\ }\href@noop {} {\bibfield  {journal} {\bibinfo  {journal} {Optica}\ }\textbf {\bibinfo {volume} {9}},\ \bibinfo {pages} {724} (\bibinfo {year} {2022})}\BibitemShut {NoStop}%
\bibitem [{\citenamefont {Pendry}(2023)}]{Pendry:ph}%
  \BibitemOpen
  \bibfield  {author} {\bibinfo {author} {\bibfnamefont {J.~B.}\ \bibnamefont {Pendry}},\ }\bibfield  {title} {\bibinfo {title} {Photon number conservation in time dependent systems [invited]},\ }\href {https://opg.optica.org/oe/abstract.cfm?URI=oe-31-1-452} {\bibfield  {journal} {\bibinfo  {journal} {Opt. Express}\ }\textbf {\bibinfo {volume} {31}},\ \bibinfo {pages} {452} (\bibinfo {year} {2023})}\BibitemShut {NoStop}%
\bibitem [{\citenamefont {Apffel}\ and\ \citenamefont {Fort}(2022)}]{Fort22}%
  \BibitemOpen
  \bibfield  {author} {\bibinfo {author} {\bibfnamefont {B.}~\bibnamefont {Apffel}}\ and\ \bibinfo {author} {\bibfnamefont {E.}~\bibnamefont {Fort}},\ }\bibfield  {title} {\bibinfo {title} {Frequency conversion cascade by crossing multiple space and time interfaces},\ }\href {https://doi.org/10.1103/PhysRevLett.128.064501} {\bibfield  {journal} {\bibinfo  {journal} {Phys. Rev. Lett.}\ }\textbf {\bibinfo {volume} {128}},\ \bibinfo {pages} {064501} (\bibinfo {year} {2022})}\BibitemShut {NoStop}%
\bibitem [{\citenamefont {{n}a}\ and\ \citenamefont {Engheta}(2024)}]{PachecoPena24}%
  \BibitemOpen
  \bibfield  {author} {\bibinfo {author} {\bibfnamefont {V.~P.-P.}\ \bibnamefont {{n}a}}\ and\ \bibinfo {author} {\bibfnamefont {N.}~\bibnamefont {Engheta}},\ }\bibfield  {title} {\bibinfo {title} {Spatiotemporal cascading of dielectric waveguides [invited]},\ }\href {https://doi.org/10.1364/OME.516262} {\bibfield  {journal} {\bibinfo  {journal} {Opt. Mater. Express}\ }\textbf {\bibinfo {volume} {14}},\ \bibinfo {pages} {1062} (\bibinfo {year} {2024})}\BibitemShut {NoStop}%
\bibitem [{\citenamefont {Zhang}\ \emph {et~al.}(2024)\citenamefont {Zhang}, \citenamefont {Donaldson},\ and\ \citenamefont {Agrawal}}]{PhysRevA.110.043526}%
  \BibitemOpen
  \bibfield  {author} {\bibinfo {author} {\bibfnamefont {J.}~\bibnamefont {Zhang}}, \bibinfo {author} {\bibfnamefont {W.}~\bibnamefont {Donaldson}},\ and\ \bibinfo {author} {\bibfnamefont {G.~P.}\ \bibnamefont {Agrawal}},\ }\bibfield  {title} {\bibinfo {title} {Conservation law for electromagnetic fields in a space-time-varying medium and its implications},\ }\href@noop {} {\bibfield  {journal} {\bibinfo  {journal} {Phys. Rev. A}\ }\textbf {\bibinfo {volume} {110}},\ \bibinfo {pages} {043526} (\bibinfo {year} {2024})}\BibitemShut {NoStop}%
\bibitem [{\citenamefont {Liberal}\ \emph {et~al.}(2024)\citenamefont {Liberal}, \citenamefont {Ganfornina-Andrades},\ and\ \citenamefont {Vázquez-Lozano}}]{liberal:Noether}%
  \BibitemOpen
  \bibfield  {author} {\bibinfo {author} {\bibfnamefont {I.}~\bibnamefont {Liberal}}, \bibinfo {author} {\bibfnamefont {A.}~\bibnamefont {Ganfornina-Andrades}},\ and\ \bibinfo {author} {\bibfnamefont {J.~E.}\ \bibnamefont {Vázquez-Lozano}},\ }\bibfield  {title} {\bibinfo {title} {Spatiotemporal symmetries and energy-momentum conservation in uniform spacetime metamaterials},\ }\href@noop {} {\bibfield  {journal} {\bibinfo  {journal} {ACS Photonics}\ } (\bibinfo {year} {2024})}\BibitemShut {NoStop}%
\bibitem [{\citenamefont {Bahrami}\ \emph {et~al.}(2025)\citenamefont {Bahrami}, \citenamefont {Kinder}, \citenamefont {Li},\ and\ \citenamefont {Caloz}}]{Wedges}%
  \BibitemOpen
  \bibfield  {author} {\bibinfo {author} {\bibfnamefont {A.}~\bibnamefont {Bahrami}}, \bibinfo {author} {\bibfnamefont {K.~D.}\ \bibnamefont {Kinder}}, \bibinfo {author} {\bibfnamefont {Z.}~\bibnamefont {Li}},\ and\ \bibinfo {author} {\bibfnamefont {C.}~\bibnamefont {Caloz}},\ }\bibfield  {title} {\bibinfo {title} {Space-time wedges},\ }\href {https://doi.org/10.1515/nanoph-2024-0526} {\bibfield  {journal} {\bibinfo  {journal} {Nanophotonics}\ } (\bibinfo {year} {2025})}\BibitemShut {NoStop}%
\bibitem [{\citenamefont {Li}\ \emph {et~al.}(2024)\citenamefont {Li}, \citenamefont {Ma}, \citenamefont {Deck-Léger}, \citenamefont {Bahrami},\ and\ \citenamefont {Caloz}}]{Caloz2024}%
  \BibitemOpen
  \bibfield  {author} {\bibinfo {author} {\bibfnamefont {Z.}~\bibnamefont {Li}}, \bibinfo {author} {\bibfnamefont {X.}~\bibnamefont {Ma}}, \bibinfo {author} {\bibfnamefont {Z.-L.}\ \bibnamefont {Deck-Léger}}, \bibinfo {author} {\bibfnamefont {A.}~\bibnamefont {Bahrami}},\ and\ \bibinfo {author} {\bibfnamefont {C.}~\bibnamefont {Caloz}},\ }\bibfield  {title} {\bibinfo {title} {Wave-medium interactions in dynamic matter and modulation systems},\ }\href@noop {} {\bibfield  {journal} {\bibinfo  {journal} {arXiv:2404.00079}\ } (\bibinfo {year} {2024})}\BibitemShut {NoStop}%
\bibitem [{\citenamefont {Hadad}\ and\ \citenamefont {Shlivinski}(2020)}]{Yakir2020}%
  \BibitemOpen
  \bibfield  {author} {\bibinfo {author} {\bibfnamefont {Y.}~\bibnamefont {Hadad}}\ and\ \bibinfo {author} {\bibfnamefont {A.}~\bibnamefont {Shlivinski}},\ }\bibfield  {title} {\bibinfo {title} {Soft temporal switching of transmission line parameters: Wave-field, energy balance, and applications},\ }\href {https://doi.org/10.1109/TAP.2020.2967302} {\bibfield  {journal} {\bibinfo  {journal} {IEEE Transactions on Antennas and Propagation}\ }\textbf {\bibinfo {volume} {68}},\ \bibinfo {pages} {1643} (\bibinfo {year} {2020})}\BibitemShut {NoStop}%
\bibitem [{\citenamefont {Southwell}(1983)}]{gradient}%
  \BibitemOpen
  \bibfield  {author} {\bibinfo {author} {\bibfnamefont {W.~H.}\ \bibnamefont {Southwell}},\ }\bibfield  {title} {\bibinfo {title} {Gradient-index antireflection coatings},\ }\href {https://doi.org/10.1364/OL.8.000584} {\bibfield  {journal} {\bibinfo  {journal} {Opt. Lett.}\ }\textbf {\bibinfo {volume} {8}},\ \bibinfo {pages} {584} (\bibinfo {year} {1983})}\BibitemShut {NoStop}%
\bibitem [{\citenamefont {Liberal}\ \emph {et~al.}(2023)\citenamefont {Liberal}, \citenamefont {Vázquez-Lozano},\ and\ \citenamefont {Pacheco-Peña}}]{Liberal2023}%
  \BibitemOpen
  \bibfield  {author} {\bibinfo {author} {\bibfnamefont {I.}~\bibnamefont {Liberal}}, \bibinfo {author} {\bibfnamefont {J.~E.}\ \bibnamefont {Vázquez-Lozano}},\ and\ \bibinfo {author} {\bibfnamefont {V.}~\bibnamefont {Pacheco-Peña}},\ }\bibfield  {title} {\bibinfo {title} {Quantum antireflection temporal coatings: Quantum state frequency shifting and inhibited thermal noise amplification},\ }\href@noop {} {\bibfield  {journal} {\bibinfo  {journal} {Laser \& Photonics Reviews}\ }\textbf {\bibinfo {volume} {17}},\ \bibinfo {pages} {1} (\bibinfo {year} {2023})}\BibitemShut {NoStop}%
\bibitem [{\citenamefont {Fox}(2006)}]{quantum.optics}%
  \BibitemOpen
  \bibfield  {author} {\bibinfo {author} {\bibfnamefont {M.}~\bibnamefont {Fox}},\ }\href@noop {} {\emph {\bibinfo {title} {Quantum Optics: An Introduction}}}\ (\bibinfo  {publisher} {Oxford University Press},\ \bibinfo {year} {2006})\BibitemShut {NoStop}%
\bibitem [{SM()}]{SM}%
  \BibitemOpen
  \href@noop {} {}\bibinfo {note} {Supplementary Materials}\BibitemShut {NoStop}%
\bibitem [{\citenamefont {Longair}(2011)}]{longair2011high}%
  \BibitemOpen
  \bibfield  {author} {\bibinfo {author} {\bibfnamefont {M.~S.}\ \bibnamefont {Longair}},\ }\href@noop {} {\emph {\bibinfo {title} {High-Energy Astrophysics}}},\ \bibinfo {edition} {3rd}\ ed.\ (\bibinfo  {publisher} {Cambridge University Press},\ \bibinfo {address} {Cambridge},\ \bibinfo {year} {2011})\BibitemShut {NoStop}%
\bibitem [{\citenamefont {Mirmoosa}\ \emph {et~al.}(2024)\citenamefont {Mirmoosa}, \citenamefont {Mostafa}, \citenamefont {Norrman},\ and\ \citenamefont {Tretyakov}}]{Sajjad2024}%
  \BibitemOpen
  \bibfield  {author} {\bibinfo {author} {\bibfnamefont {M.~S.}\ \bibnamefont {Mirmoosa}}, \bibinfo {author} {\bibfnamefont {M.~H.}\ \bibnamefont {Mostafa}}, \bibinfo {author} {\bibfnamefont {A.}~\bibnamefont {Norrman}},\ and\ \bibinfo {author} {\bibfnamefont {S.~A.}\ \bibnamefont {Tretyakov}},\ }\bibfield  {title} {\bibinfo {title} {Time interfaces in bianisotropic media},\ }\href {https://doi.org/10.1103/PhysRevResearch.6.013334} {\bibfield  {journal} {\bibinfo  {journal} {Phys. Rev. Res.}\ }\textbf {\bibinfo {volume} {6}},\ \bibinfo {pages} {013334} (\bibinfo {year} {2024})}\BibitemShut {NoStop}%
\bibitem [{\citenamefont {Pacheco-Pe{\~n}a}\ and\ \citenamefont {Engheta}(2020)}]{pacheco2020anti}%
  \BibitemOpen
  \bibfield  {author} {\bibinfo {author} {\bibfnamefont {V.}~\bibnamefont {Pacheco-Pe{\~n}a}}\ and\ \bibinfo {author} {\bibfnamefont {N.}~\bibnamefont {Engheta}},\ }\bibfield  {title} {\bibinfo {title} {Antireflection temporal coatings},\ }\href@noop {} {\bibfield  {journal} {\bibinfo  {journal} {Optica}\ }\textbf {\bibinfo {volume} {7}},\ \bibinfo {pages} {323} (\bibinfo {year} {2020})}\BibitemShut {NoStop}%
\end{thebibliography}%

\end{document}